\documentclass{ws-procs975x65}

\newcommand{\bwt}{\begin{widetext}}
\newcommand{\ewt}{\end{widetext}}
\newcommand{\beq}{\begin{equation}}
\newcommand{\eeq}{\end{equation}}
\newcommand{\bea}{\begin{eqnarray}}
\newcommand{\eea}{\end{eqnarray}}

\begin{document}

\title{The Equation of State and Cooling of Hyperonic Neutron Stars}

\author{Laura Tolos$^*$} 

\address{Institut f\"ur Theoretische Physik, Goethe Universit\"at Frankfurt, \\Max von Laue Strasse 1, 60438 Frankfurt, Germany\\
Frankfurt Institute for Advanced Studies,  Goethe Universit\"at Frankfurt, \\ Ruth-Moufang-Str. 1, 60438 Frankfurt am Main, Germany \\
Institute of Space Sciences (ICE, CSIC), Campus UAB, \\ Carrer de Can Magrans, 08193, Barcelona, Spain \\
Institut d'Estudis Espacials de Catalunya (IEEC), 08034 Barcelona, Spain \\
$^*$E-mail: tolos@th.physik.uni-frankfurt.de\\
}

\author{Mario Centelles and Angels Ramos} 

\address{Departament de F\'{\i}sica Qu\`antica i Astrof\'{\i}sica and Institut de Ci\`encies del Cosmos
(ICCUB), Facultat de F\'{\i}sica, Universitat de Barcelona, Mart\'{\i} i Franqu\`es 1, 08028
Barcelona, Spain}

\author{Rodrigo Negreiros}
\address{Instituto de F\'isica, Universidade Federal Fluminense, \\ Av. Gal. Milton Tavares S/N, Niter\'oi, Brazil}

\author{Veronica Dexheimer}
\address{Department of Physics, Kent State University, Kent OH 44242 USA}

\begin{abstract}
We present two recent parametrizations of the equation of state (FSU2R and FSU2H models) that reproduce the properties of nuclear matter and finite nuclei, fulfill constraints on high-density matter stemming from heavy-ion collisions, produce 2$M_{\odot}$ neutron stars, and generate neutron star radii below 13 km. Making use of these equations of state, cooling simulations for isolated neutron stars are performed.  We find that two of the models studied, FSU2R (with nucleons) and, in particular, FSU2H (with nucleons and hyperons), show very good agreement with cooling observations, even without including nucleon pairing. This indicates that cooling observations are compatible with an equation of state that produces a soft nuclear symmetry energy and, thus, generates small neutron star radii.  Nevertheless, both schemes produce cold isolated neutron stars with masses above $1.8 M_{\odot}$.
\end{abstract}


\bodymatter


\section{Introduction}

The thermal evolution (or cooling history) of neutron stars strongly depends on the equation of state (EoS) and the associated composition of these compact objects  \citep{Page2006,Weber2007}. Several works have addressed the cooling history of neutron stars and they agree that it is imperative to establish whether 
 fast cooling processes take place, since if the star cools down too fast, it will yield to a disagreement with most observed data.  If the stellar proton fraction is high enough, the most important fast processes  (the so-called direct URCA, DU)  will take place, thus leading to a direct connection between the thermal behavior and the symmetry energy of nuclear matter.

In this paper we investigate the thermal evolution of neutrons stars using  EoSs for the nucleonic and hyperonic inner core \cite{Tolos:2016hhl,Tolos:2017lgv} that reconcile the $2 M_{\odot}$ mass observations  
\citep{Demorest:2010bx,Antoniadis:2013pzd} with determinations of stellar radii below 13~km  
(see Ref.~\refcite{Fortin:2014mya} for an overview), the latter being confirmed from  analysis of the gravitational-wave emission of GW170817 (see Table II in Ref.~\refcite{Montana:2018bkb} and references therein) detected by the LIGO and Virgo collaborations \citep{TheLIGOScientific:2017qsa}. Moreover, the aforementioned microscopic models satisfy the properties of nuclear matter and finite nuclei  \citep{Tsang:2012se,Chen:2014sca} and constraints from heavy-ion collisions (HICs) \citep{Danielewicz:2002pu,Fuchs:2000kp,Lynch:2009vc}.

In particular, the two models FSU2R and FSU2H, based on the nucleonic FSU2 model \cite{Chen:2014sca},  differ on the onset of appearance of each hyperon, whereas the neutron star maximum masses calculated with these models show only a moderate dispersion of about~0.1$M_{\odot}$ \citep{Tolos:2017lgv}.  In the present study, we focus on how the hyperons, as well as the symmetry energy of the microscopic model influence the cooling history of isolated neutron stars, considering also different nucleon pairing  scenarios \cite{Negreiros:2018cho}.

\section{FSU2R and FSU2H models}

\begin{figure}[t!]
\begin{center}
\includegraphics[width=0.47\columnwidth,height=0.36\columnwidth]{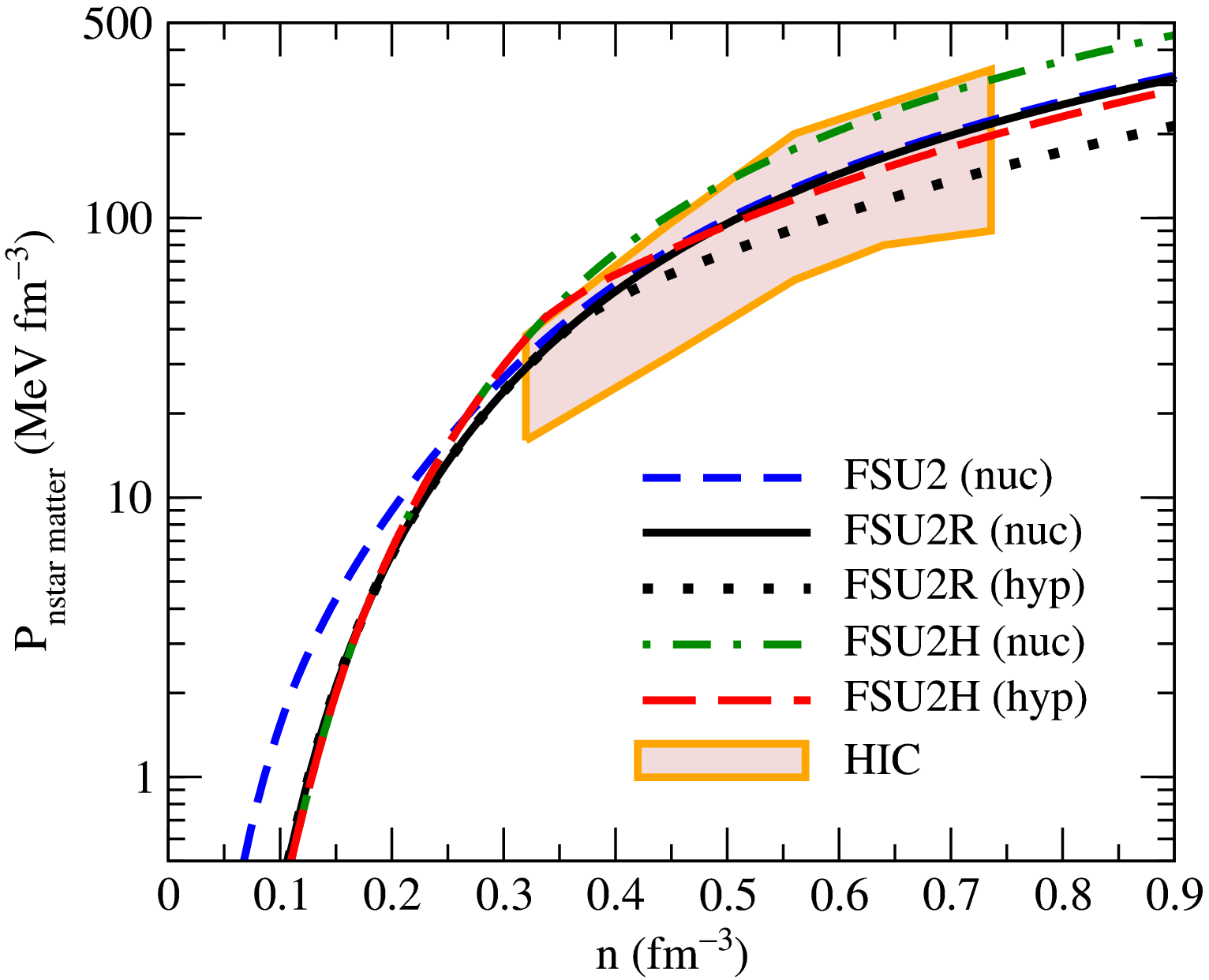}
\hfill
\includegraphics[width=0.47\textwidth,height=0.36\columnwidth]{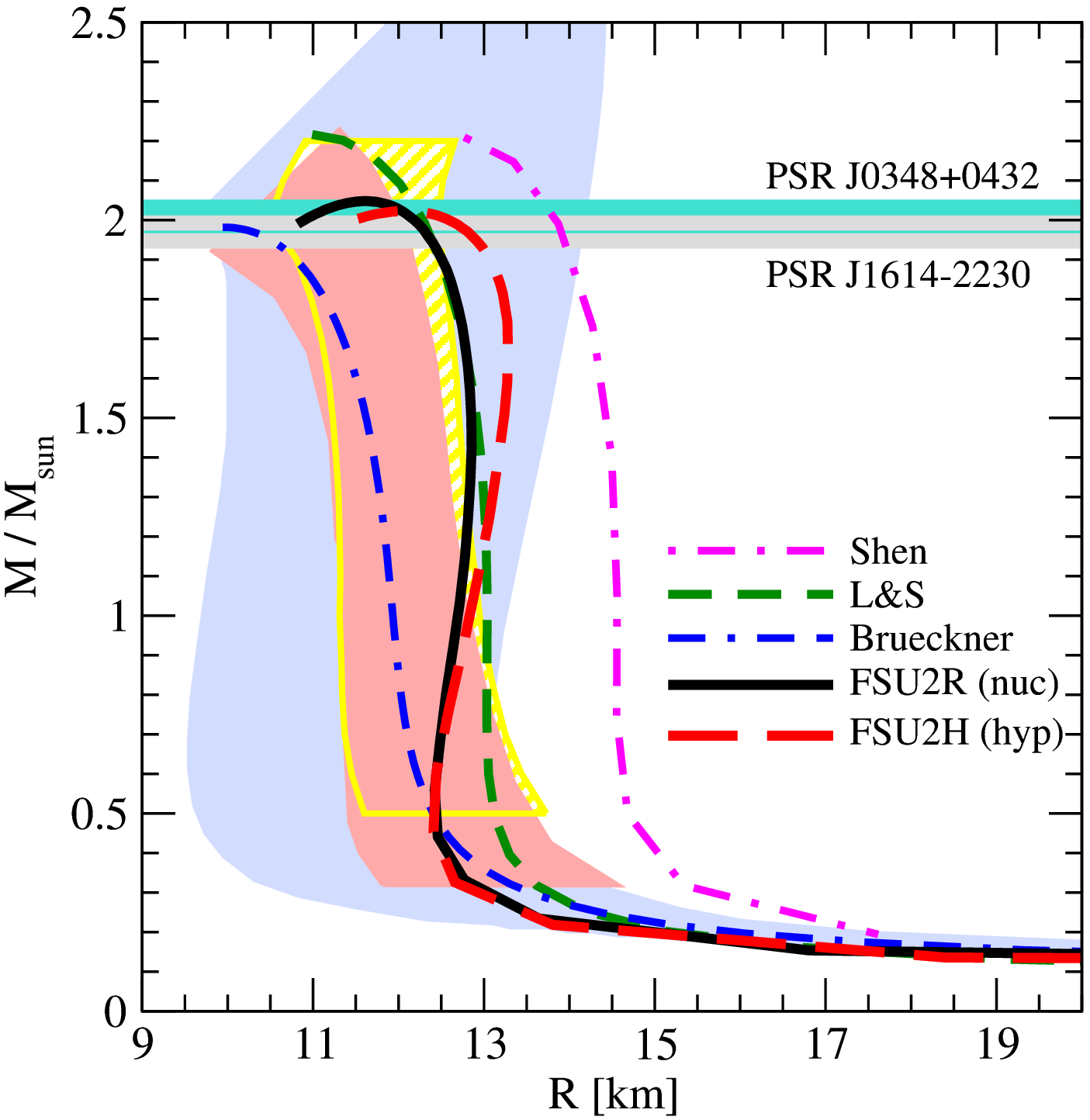}
\caption{Left: Pressure of $\beta$-stable neutron star matter as a function of baryon number density for different models \cite{Negreiros:2018cho}, with the colored area being compatible with HiCs \citep{Danielewicz:2002pu}.  Right: Mass versus radius for neutron stars from the models FSU2R and FSU2H  \cite{Tolos:2017lgv} together with Shen \citep{Shen:1998gq}, L\&S \citep{ls91}, Brueckner \citep{Sharma:2015bna}. The thin horizontal bands indicate the $2 M_{\odot}$ observations  \citep{Demorest:2010bx,Antoniadis:2013pzd}, the vertical blue band  is the region constrained from chiral nuclear interaction \citep{Hebeler:2013nza}, the vertical red band is the area derived from five quiescent low-mass
X-ray binaries and five photospheric radius expansion X-ray bursters  \citep{Lattimer:2014sga}, and the vertical striped yellow band is the mass-radius constraint from cooling tails of type-I
X-ray bursts in three low-mass X-ray binaries \citep{Nattila:2015jra}. }
\label{fig:eos}
\end{center}
\end{figure}

Our models are based on two new parametrizations of the FSU2 RMF model \citep{Chen:2014sca}. We start by considering only nucleons. The FSU2R(nuc) model produces a soft symmetry energy and a soft pressure of neutron matter for densities $n \lesssim 2n_0$, as seen in Fig.~\ref{fig:eos} (left). This is done by increasing the $\Lambda_{\omega}$ coefficient of the mixed quartic isovector-vector interaction that modifies the density dependence of the nuclear symmetry energy, thus turning the EoS softer \citep{Tolos:2016hhl,Tolos:2017lgv}. As a result, radii within the range of 11.5--13~km are obtained from FSU2R for neutron stars with masses between the maximum mass and $M= 1.4 M_\odot$, as seen in Fig.~\ref{fig:eos} (right). FSU2R 
predicts $E_{\rm sym}(n_0)= 30.7$ MeV and $L=46.9$ MeV \citep{Tolos:2017lgv}, that differs from the FSU2 model with $E_{\rm sym}(n_0)= 37.6$ MeV and $L=112.8$ MeV and is in agreement with the limits of recent determinations (see Fig.1 of Ref.~\refcite{Tolos:2017lgv} for a summary and references therein). In the high-density sector of the EoS, the FSU2R and FSU2 EoSs are similar, and, hence, FSU2R also reproduces heavy neutron stars (see Fig.~\ref{fig:eos} (right)). 

When hyperons are incorporated, a softening of the high-density EoS with hyperonic degrees of freedom takes place (compare the FSU2R(hyp) and 
FSU2R(nuc) EoSs in Fig.~\ref{fig:eos} (left)), so we obtain a reduction of the maximal neutron star mass below 2$M_{\odot}$ in FSU2R(hyp). We may readjust  the parameters of the nuclear model by stiffening the EoS of isospin-symmetric matter for densities above twice the saturation density, where hyperons set in. We do that by changing the quartic isoscalar-vector self-interaction (with reducing coupling $\zeta$), which stiffens the EoS at high densities \citep{Tolos:2016hhl,Tolos:2017lgv}. The FSU2H allowing for hyperons, FSU2H(hyp), successfully fulfills the 2$M_{\odot}$ mass limit 
with moderate radii for the star (see Fig.~\ref{fig:eos} (right)), while the base nuclear model  still reproduces the 
properties of nuclear matter and nuclei, with $E_{\rm sym}(n_0)= 30.5$ MeV and $L=44.5$ MeV \citep{Tolos:2017lgv}. The parameter set of FSU2R(nuc) and FSU2H(hyp) and some nuclear matter properties can be found in Refs.~\refcite{Tolos:2017lgv,Negreiros:2018cho}.

\section{Cooling of neutron stars}

\begin{figure}[t]
\begin{center}
\includegraphics[width=0.47\columnwidth]{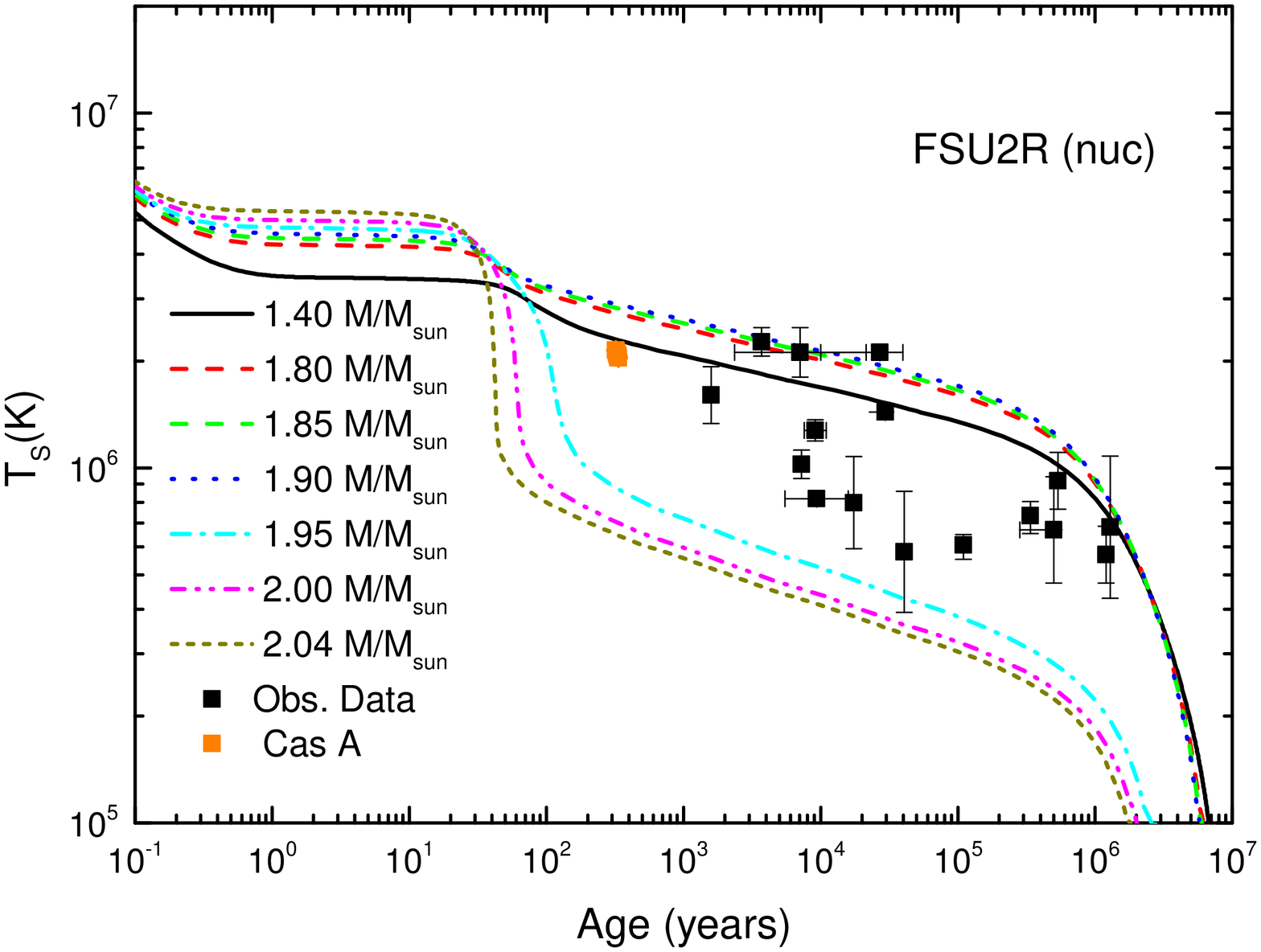}
\hfill
\includegraphics[width=0.47\columnwidth]{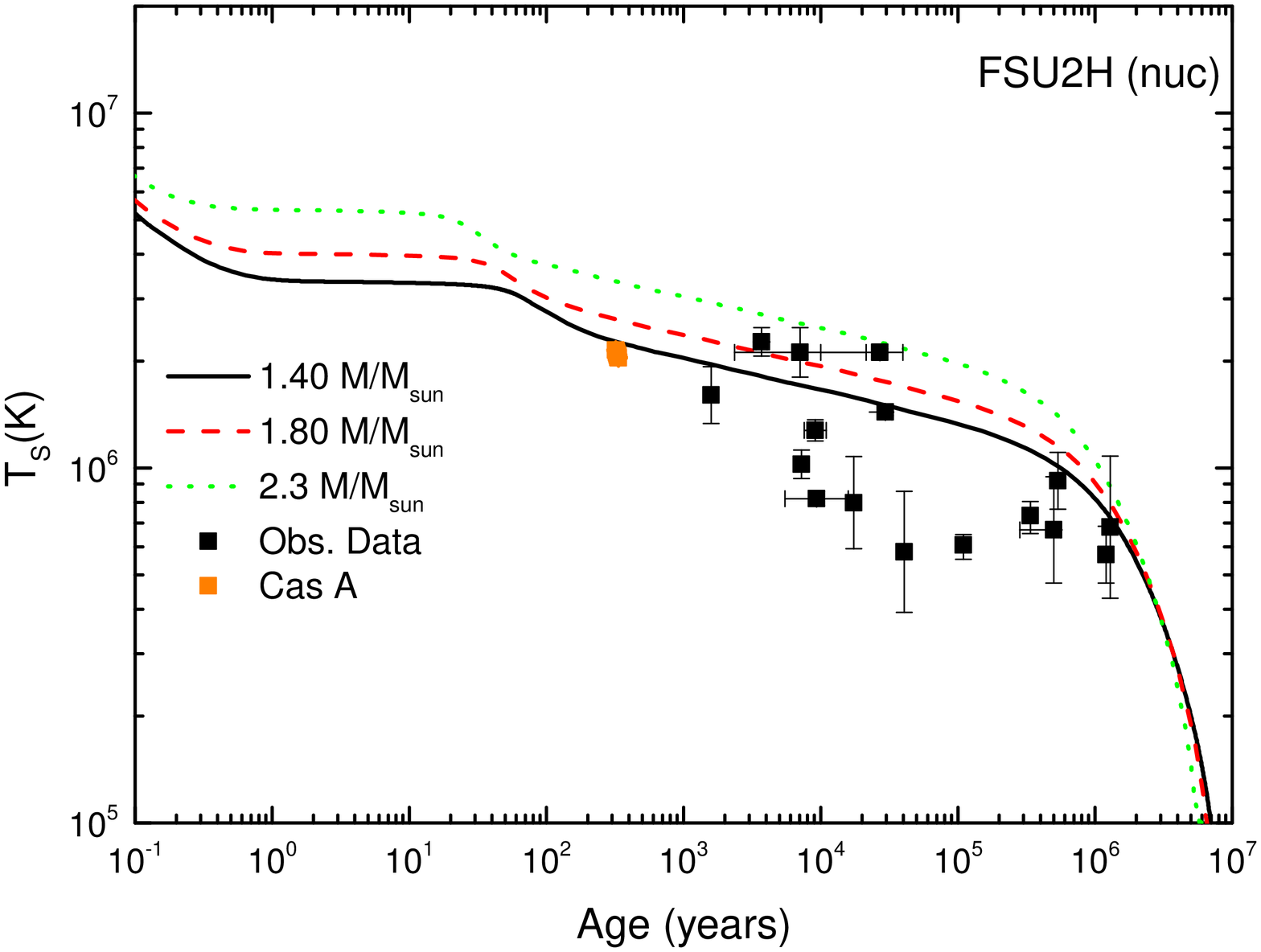}
\caption{Surface temperature as a function of the stellar age for different 
neutron star masses for FSU2R(nuc) (left) and FSU2H(nuc) (right). Also shown are different observed thermal 
data. \cite{Negreiros:2018cho}}
\label{fig:fsu2r}
\end{center}
\end{figure}

\begin{figure}[t]
\begin{center}
\includegraphics[width=0.47\columnwidth]{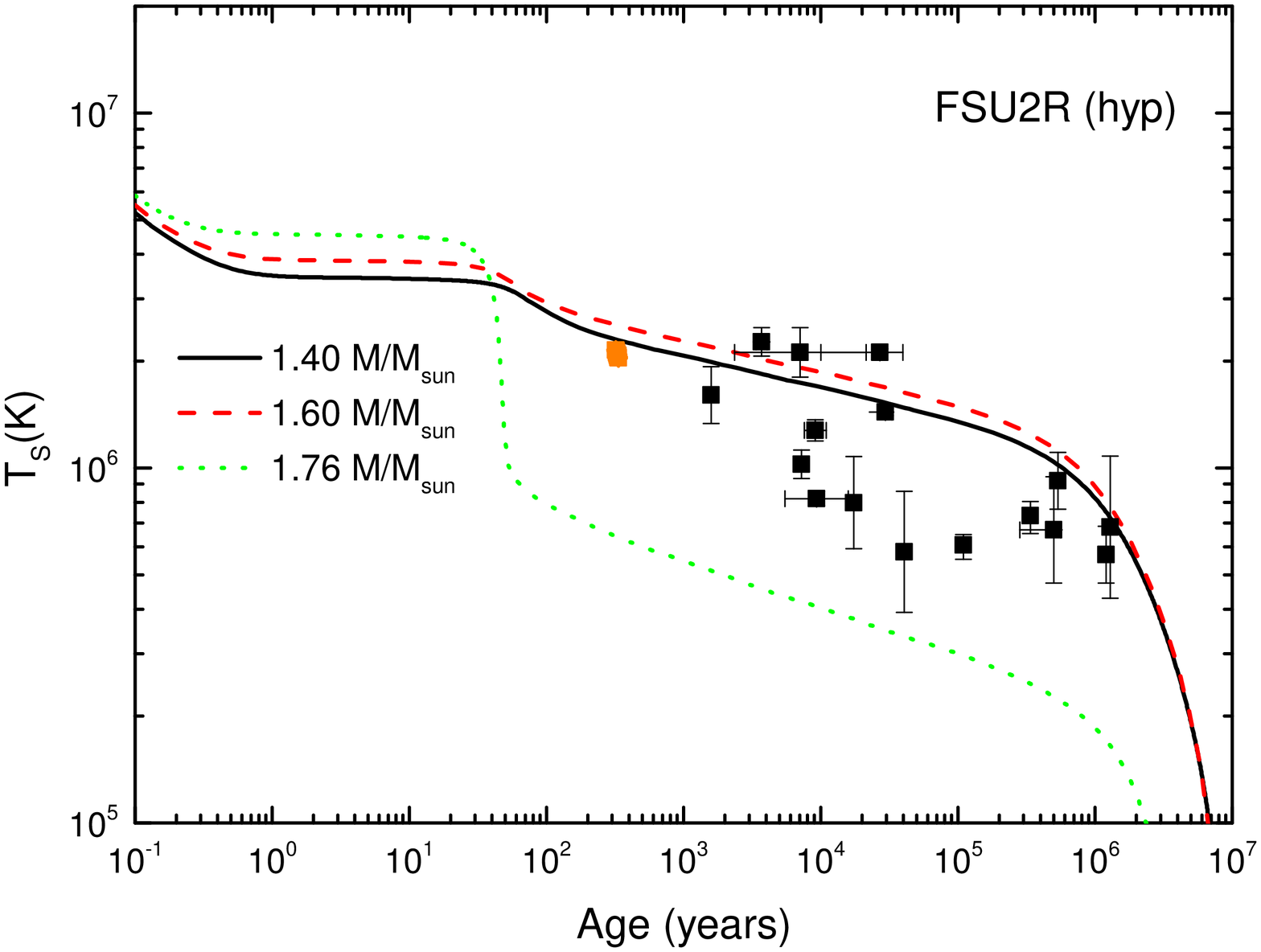}
\hfill
\includegraphics[width=0.47\columnwidth]{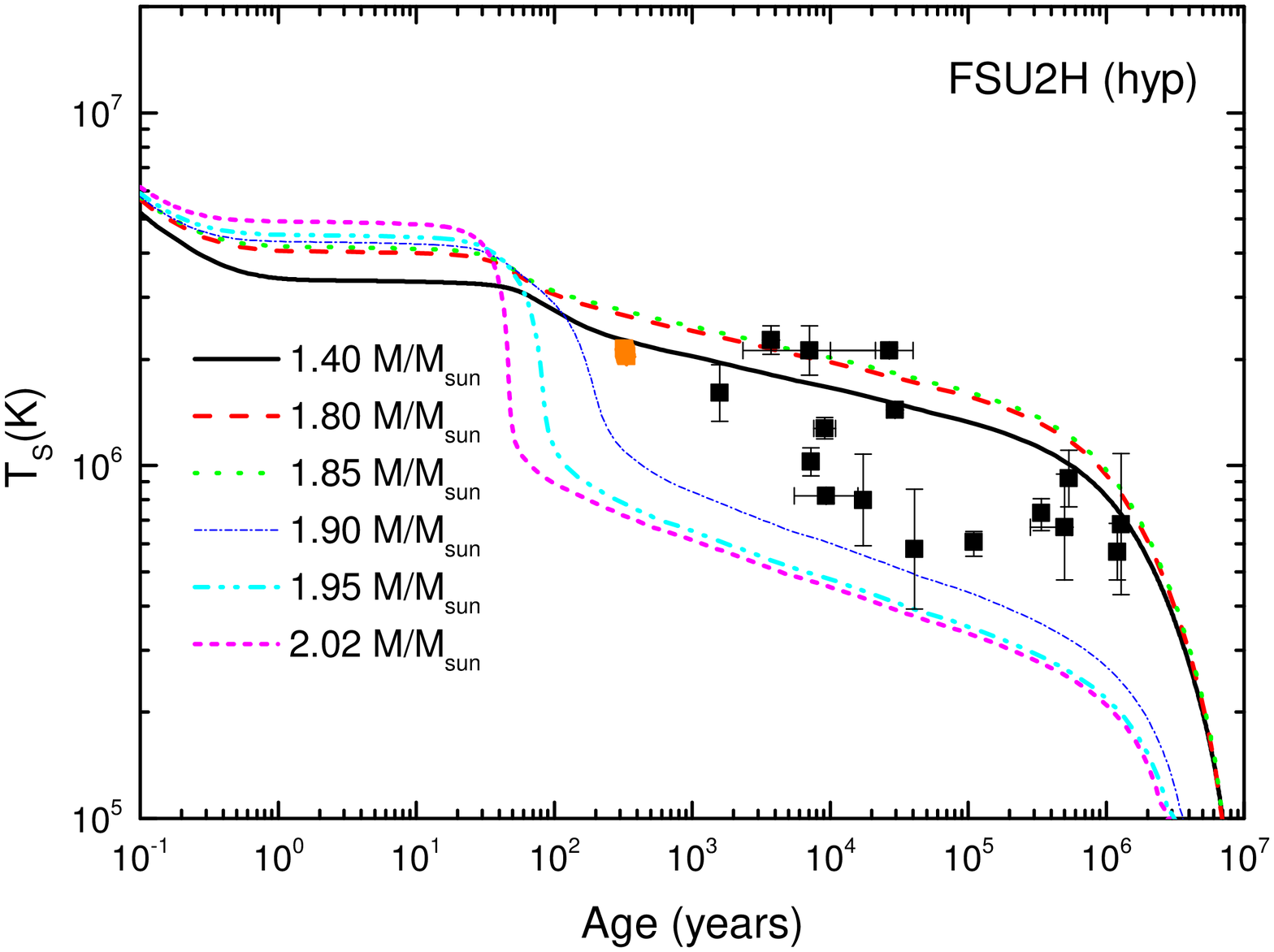}
\caption{Surface temperature as a function of the stellar age for different 
neutron star masses for FSU2R(hyp) (left) and FSU2H(hyp) (right). Also shown are different observed thermal 
data.\cite{Negreiros:2018cho} }
\label{fig:fsu2h}
\end{center}
\end{figure}

We first consider the thermal evolution of neutron stars without taking into account any sort of pairing 
(neither in the core nor in the crust). This is done in  Figs.~\ref{fig:fsu2r} and \ref{fig:fsu2h} for the FSU2R and FSU2H models, with and without hyperons. The figures also display the observed surface temperature versus the age of a set of prominent neutron stars, including that of the remnant in Cas A. The nucleonic DU and hyperonic DU thresholds from each microscopic model are given in Table~\ref{table:cool1} for low-mass to high-mass neutron stars. When DU reactions  are allowed in the model, that is, the DU threshold takes place for densities below the central density of the star, they lead to an enhanced cooling of the star, thus to a fast cooling. 

Several conclusions can be drawn from the comparison between the different plots in Figs.~\ref{fig:fsu2r} and \ref{fig:fsu2h}  together with the analysis of Table~\ref{table:cool1}:

\begin{itemize}
\item Low-mass stars of $1.4 M_{\odot}$ when only nucleons are considered: the cooling pattern of a $1.4 M_{\odot}$ star is slow for FSU2R(nuc) and FSU2H(nuc). This is mainly due to the density dependence of the symmetry energy around saturation and, hence, to the symmetry energy slope parameter ($L$). The 
smaller the value of $L$ is, the less protons are produced and, thus, the DU process appears at higher densities, making the cooling less efficient.
\item High-mass stars of $1.8-2 M_{\odot}$ when only nucleons are considered:  the  different behaviours exhibited by the cooling curves of FSU2R(nuc) (fast cooling) and 
FSU2H(nuc) (slow cooling) are correlated with the different values of the central densities in these stars, as seen in Table~\ref{table:cool1}. The stiffer the EoS is, the lower central densities are and the slower the cooling is.
\item The inclusion of hyperons in medium- to heavy-mass stars speeds up the cooling (compare  in Figs.~\ref{fig:fsu2r} and \ref{fig:fsu2h} FSU2R(nuc) and FSU2R(hyp) or FSU2H(nuc) and FSU2H(hyp)).  On the one hand, this is  because the presence of hyperons (mostly $\Lambda$ particles) reduces the neutron fraction at a given baryon number density and, consequently, the DU restriction 
$\vec{k}_{Fn}=\vec{k}_{Fp}+\vec{k}_{Fe}$,  can be fulfilled at a lower density. Here ${k}_{Fn}$, ${k}_{Fp}$ and ${k}_{Fe}$ are the Fermi momenta of the neutron, proton and electron, respectively. On the other hand, the appearance of hyperons softens the EoS, thus FSU2R(hyp) and FSU2H(hyp) produce stars with higher central densities than the nucleonic counterparts, so that they may overcome the DU threshold (see Table~\ref{table:cool1}).
\end{itemize}

\begin{table}[t]
\tbl{The nucleonic DU thresholds for FSU2(nuc), FSU2R(nuc), FSU2R(hyp), FSU2H(nuc) and FSU2H(hyp) models, together with the hyperonic DU threshold for each of them. Also shown is the central density $n_c$ for three selected neutron star masses. \citep{Negreiros:2018cho} }
{\begin{tabular}{| c c c c c c |}
\hline
Models & $DU_{th}$ & hyp $DU_{th}$&  $1.4 M_\odot$   & $1.76 M_\odot$     &$ 2.0 M_\odot$  \\
&(fm$^{-3}$) &(fm$^{-3}$) & $n_c$ (fm$^{-3}$)  & $n_c$ (fm$^{-3}$)  & $n_c$ (fm$^{-3}$)    \\
\hline
FSU2 (nuc)  & 0.21 & | & 0.35 &  0.47 &  0.64 \\
FSU2R (nuc) & 0.61 & |    & 0.39 &  0.51 &  0.72  \\
FSU2H (nuc) & {0.61} & |    & 0.34 &  0.39 &  0.45  \\
FSU2R (hyp) & 0.57 & 0.37 & 0.40 &  0.87 &  |    \\
FSU2H (hyp) & 0.52 & 0.34 & 0.34 &  0.44 &  0.71 \\
\hline
 \end{tabular}}
 \label{table:cool1}
\end{table}

We now  investigate the effect of neutron superfluidity and proton superconductivity on the cooling of neutron stars. We study the two most relevant cases, the FSU2R(nuc) and the FSU2H(hyp) models, which are the ones that best reproduce the observed data on cooling in Figs.~\ref{fig:fsu2r}  and \ref{fig:fsu2h}. On the one hand, the introduction of a superfluidity (conductivity) gap in the energy spectrum of baryons reduces the neutrino reaction rates, leading to a sharp drop of neutrino emissivity after the matter temperature drops below the pairing critical temperature. On the other hand, a new transient neutrino emission process appears, commonly known as pair breaking-formation (PBF) process,  where two quasi-baryons with similar anti-parallel momenta annihilate into a neutrino pair.

\begin{figure}[t]
\begin{center}
\includegraphics[width=0.47\columnwidth]{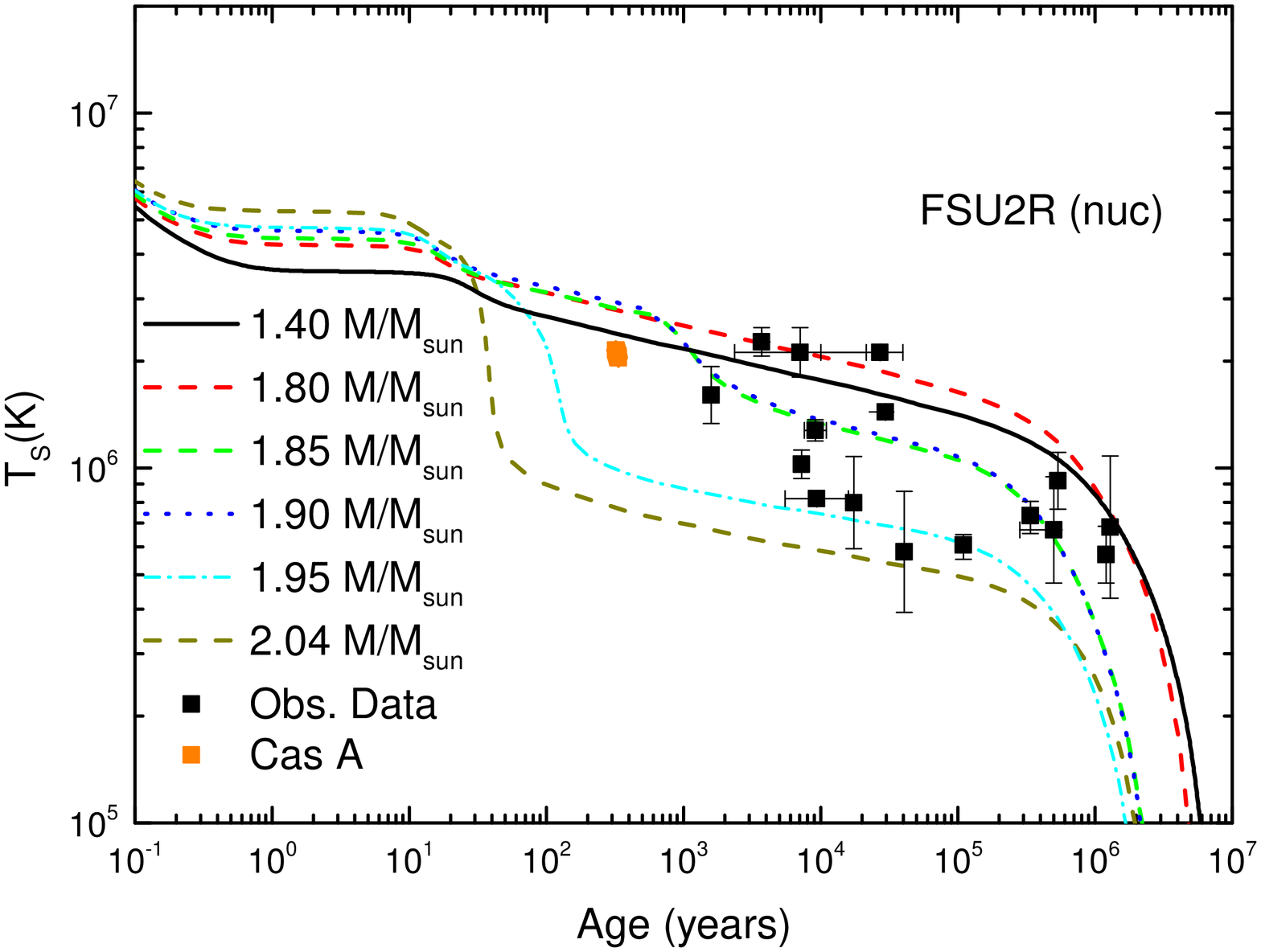}
\hfill
\includegraphics[width=0.45\columnwidth]{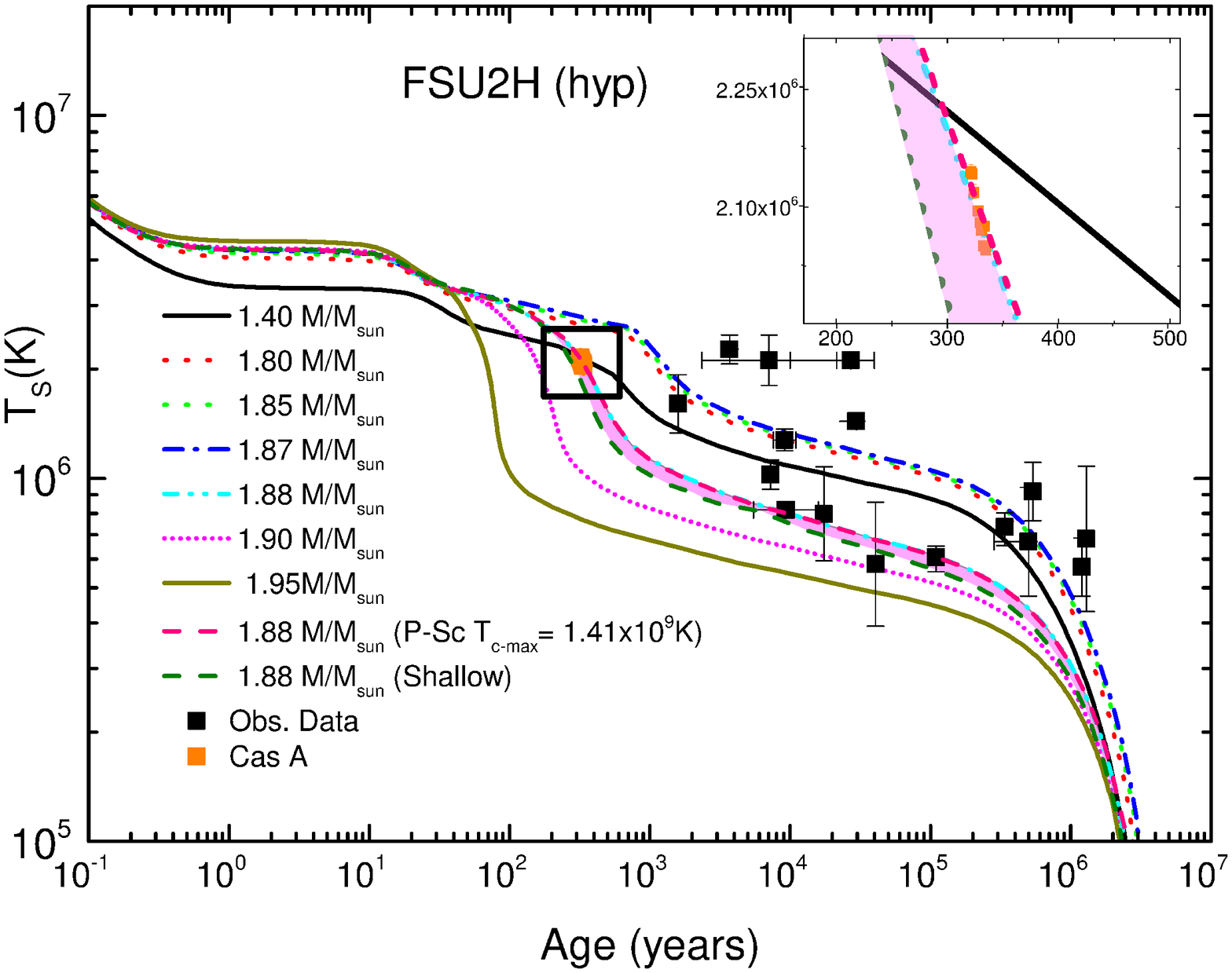}
\caption{Surface temperature as a function of the stellar age for different 
neutron star masses for FSU2R(nuc) (left) and FSU2H(hyp) (right) with  medium proton pairing and  neutron pairing. \cite{Negreiros:2018cho} }
\label{fig:pairing}
\end{center}
\end{figure}

The analysis of the plots in Fig.~\ref{fig:pairing} indicates that the inclusion of medium proton pairing, in addition to the neutron pairing, improves the agreement of the cooling curves of the FSU2R(nuc) and FSU2H(hyp) models with data, specially for Cas A in the case of the FSU2H(hyp) model, as can be seen in 
the inset of Fig.~\ref{fig:pairing} (right). Note that in a very recent publication \cite{Posselt:2018xaf}, Posselt and Pavlov have revised the Cas A cooling data, concluding that the cooling of this object could be slower. In other words, we find that a shallow/medium proton superconductivity does not lead to an over-suppression of the DU processes, a fact that, combined with miscroscopic models with a soft nuclear symmetry energy, leads to an optimum agreement with observations. However, the calculations favour rather large stellar masses to explain the observed colder stars with surface temperatures $T \lesssim 10^6$ K. 

\section*{Acknowledgements}
R.N. acknowledges support from CAPES, CNPq and INCT-FNA Proc. No. 464898/2014-5, whereas
L.T. from Grant No. FPA2016-81114-P from MINECO, Heisenberg Project Nr. 383452331 and PHAROS COST Action CA16214.
V.D. acknowledges support from NSF under grant PHY-1748621, while M.C. and A.R. from Grant No. FIS2017-87534-P and MDM-2014-0369 (ICCUB) project from MINECO.

\bibliographystyle{ws-procs975x65}
\bibliography{biblio-notitles}

\end{document}